\def\papertitle{Feature-informed Latent Space Regularization for Music Source Separation}
\def\paperauthorA{Yun-Ning Hung}
\def\paperauthorB{Alexander Lerch}

\documentclass[twoside,a4paper]{article}
\usepackage{etoolbox}
\usepackage{dafx_22a}
\usepackage{amsmath,amssymb,amsfonts,amsthm}
\usepackage{euscript}
\usepackage[T1]{fontenc}
\usepackage[utf8]{inputenc}
\usepackage{nimbusserif}
\usepackage{ifpdf}
\usepackage[english]{babel}
\usepackage{caption}
\usepackage{color}
\usepackage{subcaption}
\usepackage{units}
\usepackage{microtype}
\usepackage{paralist}
\usepackage{lineno}

\newcommand{\amy}[1]{\textcolor{black}{#1}}

\input glyphtounicode
\ninept

\newcounter{numauth}
\newcounter{listcnt}
\newcommand\authcnt[1]{\ifdefined#1 \stepcounter{numauth} \fi}

\newcommand\addauth[1]{
\ifdefined#1 
\stepcounter{listcnt}
\ifnum \value{listcnt}<\value{numauth}
\appto\authorslist{, #1}
\else
\appto\authorslist{~and~#1}
\fi
\fi}
\authcnt{\paperauthorB}
\authcnt{\paperauthorC}
\authcnt{\paperauthorD}
\authcnt{\paperauthorE}
\authcnt{\paperauthorF}
\authcnt{\paperauthorG}
\authcnt{\paperauthorH}
\authcnt{\paperauthorI}
\authcnt{\paperauthorJ}
\def\authorslist{\paperauthorA}
\addauth{\paperauthorB}
\addauth{\paperauthorC}
\addauth{\paperauthorD}
\addauth{\paperauthorE}
\addauth{\paperauthorF}
\addauth{\paperauthorG}
\addauth{\paperauthorH}
\addauth{\paperauthorI}
\addauth{\paperauthorJ}

\usepackage{times}

\newif\ifpdf
\ifx\pdfoutput\relax
\else
   \ifcase\pdfoutput
      \pdffalse
   \else
      \pdftrue
\fi

\ifpdf 
  \usepackage[pdftex,
    pdftitle={\papertitle},
    pdfauthor={\authorslist},
    pdfsubject={Proceedings of the 25th International Conference on Digital Audio Effects (DAFx20in22)},
    colorlinks=false, 
    bookmarksnumbered, 
    pdfstartview=XYZ 
  ]{hyperref}
  \usepackage[pdftex]{graphicx}
\else 
  \usepackage[dvips]{epsfig,graphicx}
  \usepackage[dvips,
    pdftitle={\papertitle},
    pdfauthor={\authorslist},
    pdfsubject={Proceedings of the 25th International Conference on Digital Audio Effects (DAFx20in22)},
    colorlinks=false, 
    bookmarksnumbered, 
    pdfstartview=XYZ 
  ]{hyperref}
\fi
\usepackage[hypcap=true]{caption}
\title{\papertitle}

\affiliation{
\paperauthorA\ and \paperauthorB\,\sthanks{\vspace{-3mm}}}
{Music Informatics Group \\ Georgia Institute of Technology \\ Atlanta, USA \\
{\tt {\{\href{mailto:yhung33@gatech.edu}{yhung33}, \href{mailto:alexander.lerch@gatech.edu}{alexander.lerch}\}}@gatech.edu}
}

\begin{document}
\ifpdf 
  \DeclareGraphicsExtensions{.png,.jpg,.pdf}
\else  
  \DeclareGraphicsExtensions{.eps}
\fi


\maketitle

\begin{abstract}
The integration of additional side information to improve music source separation has been investigated numerous times, e.g., by adding features to the input or by adding learning targets in a multi-task learning scenario. These approaches, however, require additional annotations such as musical scores, instrument labels, etc. in training and possibly during inference. The available datasets for source separation do not usually provide these additional annotations. In this work, we explore transfer learning strategies to incorporate VGGish features with a state-of-the-art source separation model; VGGish features are known to be a very condensed representation of audio content and have been successfully used in many music information retrieval tasks. We introduce three approaches to incorporate the features, including two latent space regularization methods and one naive concatenation method. Experimental results show that our proposed approaches improve several evaluation metrics for music source separation.
\end{abstract}

\section{Introduction}
\label{sec:intro}
Music source separation has been an intensively studied problem due to its numerous applications. By isolating the sound of individual instruments from a mixture of instruments, source separation systems have been used, e.g., for audio-remixing~\cite{veire2018raw}, instrument-wise equalization~\cite{adali2014source}, accompaniment generation for Karaoke systems~\cite{tachibana2016real}, or singer identification~\cite{berenzweig2002using}.

\amy{A typical music source separation pipeline often includes input representation (e.g., waveform and extracted features), the machine learning model (e.g., neural network and loss function), and the post-processing algorithm (e.g., Wiener Filtering)}. 
Although there has been some work on the input representation to improve source separation systems, e.g., investigating waveforms \cite{stoller2018wave} or complex spectrograms \cite{woosung_choi_2020_4245404} instead of the common magnitude spectrograms, most research in recent years has focused on improving source separation models through new, more powerful model architectures~\cite{woosung_choi_2020_4245404, takahashi2018mmdenselstm, stoter19}. One successful architecture is the U-net \cite{ronneberger2015u} that has been adopted and utilized for many source separation studies \cite{stoller2018wave, andreas_jansson_2017_1414934, park2018music, yuan2019skip, defossez2019music}. While the original U-net is based on a Convolutional Neural Network (CNN) and skip connections, an advanced U-net architecture proposed by Takahashi et al.\ combines the CNN with a Recurrent Neural Network (RNN) \cite{takahashi2018mmdenselstm}. \amy{Different combination of CNN, RNN, and Fully-Connected (FC) layers have been studied by Choi et al.\ \cite{woosung_choi_2020_4245404}}. The popular Open-Unmix and the more recent X-UMX models are based on a combination of RNN and FC  \cite{stoter19,sawata2020all}.

While the increasing model complexity has led to noticeable increases in the quality of the system outputs, it has also led to an increased need for computational resources particularly during training. \amy{Researchers often cannot easily meet the  GPU and memory requirements for training or inference with modern source separation systems}. This resulted in a parallel research direction aiming at improving a source separation model by adding additional information during training, inference, or both~\cite{slizovskaia2019end, hung2020multitask, manilow2020simultaneous, jansson2019joint}, which is a contrasting approach compared to increasing model complexity.

However, one drawback of these methods is that they need additional ground truth annotations for training or sometimes even during inference. The available datasets for music source separation~---most prominently the  MUSDB18 dataset~\cite{musdb18}---~do not provide these additional annotations. \amy{While generating pseudo-annotations with other machine learning models can be one solution, their usefulness depends on the correctness of the model. Furthermore, the computation time will increase with more added features.} 

In this work, we propose to leverage a feature representation learned from the model pre-trained on large-scale datasets as additional information. 
\amy{This feature representation includes more generalized features than single task-specific models}. To this end, we leverage the well-known VGGish features~\cite{45611}. The VGGish model is a very deep model trained on a very large audio dataset and provides a condensed feature representation of the input audio file. We investigate three methods to incorporate VGGish features into a state-of-the-art (SOTA) music source separation model, including novel latent space regularization methods. In summary, the main contributions of this work are:
\begin{itemize}
    \item   leveraging the information contained by mid-level features trained on a different task to improve music source separation without the need of additional ground truth annotations, and
    \item   the presentation of two novel transfer learning methods for latent space regularization.
\end{itemize}

In the following sections, we first review existing approaches on using VGGish features for music information retrieval (MIR) tasks and incorporating additional features in music source separation. Section~\ref{sec: method} provides a detailed overview of the methods we propose for our feature-informed source separation system. Section~\ref{sec: experiment} presents the evaluation results for both, a state-of-the-art (SOTA) system and our proposed system. The same section also presents a detailed analysis of the latent space after regularization.

\begin{figure*}
 \centerline{
 \includegraphics[width=0.8\textwidth]{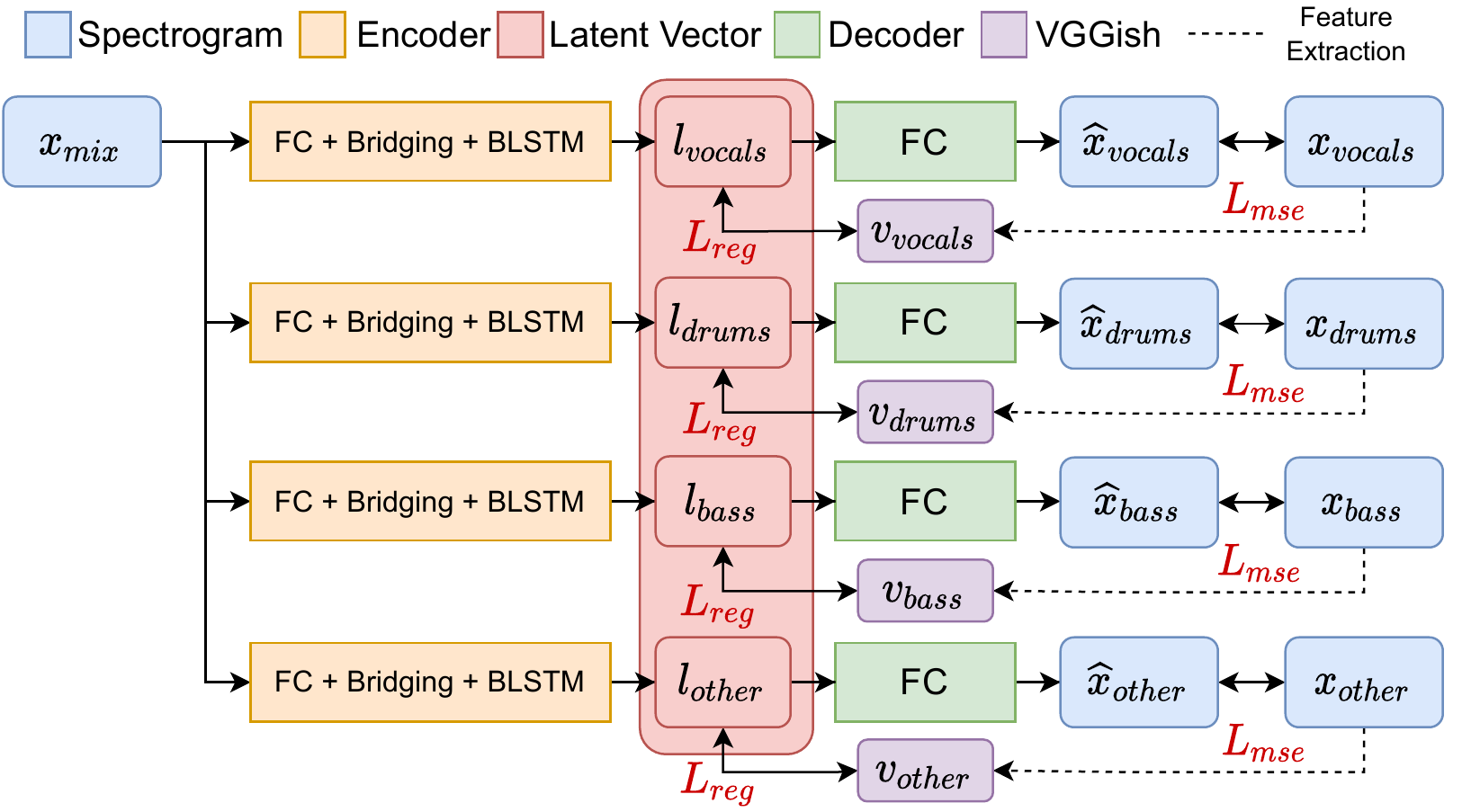}}
 \caption{Overview of the X-UMX structure and our proposed training pipeline \amy{for the two proposed regularization methods}. $\hat{x}$ denotes the estimated target spectrograms; $x$ denotes the ground truth spectrograms; $l$ denotes the latent vectors; $v$ denotes the VGGish features; $L_\mathrm{mse}$ and $L_\mathrm{reg}$ denote the Mean Square Error and Regularization loss respectively. FC denotes fully-connected layers. The dotted line represents the feature extraction process by VGGish.
 }
 \label{fig: xumx}
\end{figure*}

\section{Related Work}

\amy{This section discusses two related research directions that inspired our proposed approach. First, we introduce transfer learning methods that leverage representations from models pre-trained on larger datasets to help downstream tasks. Then, the literature on adding additional information that supports a target task is briefly surveyed.}

\subsection{VGGish features}
Traditional hand-crafted features have nowadays been replaced by learning feature representations automatically from the training data. These learned representations can also be used in other tasks in various domains. The pre-trained features from the BERT model~\cite{devlin2018bert}, for example, have been successfully used in multiple natural language processing tasks such as question answering and language inference. In the image domain, large-scale pre-trained models such as AlexNet~\cite{krizhevsky2012imagenet} and VGG-16 \cite{simonyan2014very} have achieved competitive results in image classification and assist in many visual feature extraction tasks~\cite{long2015fully}. There exist several pre-trained models in the audio domain, such as L$^3$-Net~\cite{cramer2019look}, VGGish~\cite{45611} and SoundNet \cite{aytar2016soundnet}, which leverage both audio and visual information provided by video to train an audio feature extraction network. 

\amy{VGGish features, particularly, have been successfully used in numerous MIR tasks. These tasks include weakly-supervised instrument recognition~\cite{gururani2019attention}, cross-modal representation learning~\cite{yu2019deep, zeng2018audio}, music auto-tagging~\cite{9439825}, music emotion recognition ~\cite{koh2021comparison}, and music genre classification~\cite{ramirez2020machine, choi2017transfer,gemmeke2017audio}. The VGGish model was pre-trained on a larger dataset (i.e., YouTube-8M~\cite{abu2016youtube}) than most other audio feature extractors, so it can potentially extract representations with higher discriminative power. }

\amy{The apparent popularity of VGGish features combined with the variety of the tasks they have successfully used this input representation implies that these features are able to capture many task-agnostic properties of audio files suitable for a large variety of music-related tasks. As as result, we choose the VGGish model to extract the  feature representation for our experiment. }



\subsection{Additional features for music source separation}
Leveraging additional side information to improve music source separation systems has been proposed in multiple forms. The two most common methods are 
\begin{inparaenum}[(i)]
    \item   to add the additional information directly to the input and 
    \item   to utilize the information to add a task in a multi-task setup.
\end{inparaenum}

Taking advantage of the close relation of the music instrument classification task and music source separation, several studies have shown that instrument activity labels as an additional input to a source separation system can improve results. Slizovskaia et al.\ investigated several methods to add instrument labels to the U-net model \cite{slizovskaia2020conditioned} and Swaminathan investigated adding voice activity labels to improve singing voice separation quality \cite{swaminathan_improving_2019}. Instrument activity labels can also be used as a condition to control the output sources by using one model instead of multiple models for each source \cite{slizovskaia2019end, seetharaman2019class, kadandale2020multi}. 
In addition to instrument labels, the instrumentation and pitch information provided by musical scores has also been used to guide the learning process and improve separation results \cite{ewert2014score, miron2016score, miron2017monaural}. Carabias-Orti et al.\ learn a timbre model for each instrument they separate and use the trained models as priors for the separation system based on non-negative matrix factorization \cite{carabias2013nonnegative}. 
Source separation systems can also be informed by incorporating visual features, leveraging the cross-modal information of the video for the separation \cite{zhao2018sound, yang2020remixing, gan2020music}. 

As an alternative to adding additional information to the input of a source separation system, this information can also be utilized during the training to improve the internal representation and help the model to generalize better.
Hung et al.\ proposed to combine the training of a frame-level instrument classifier and a source separation system in a multi-task setup and then leverage the instrument predictions during inference for post-processing the result \cite{hung2020multitask}.
Manilow et al.\ were able to improve source separation with a deep clustering model using both separation and transcription as a training tasks \cite{manilow2020simultaneous} and Jansson et al.\ explored a variety of methods to learn singing voice separation and fundamental frequency (F0) estimation at the same time \cite{jansson2019joint}.

\section{Proposed Method}
\label{sec: method}
In this work, we investigate three transfer learning approaches to improve the SOTA source separation model via VGGish features. The first approach directly integrates the VGGish information while the second and third approach perform transfer learning through the regularization of the latent space based on the VGGish information. Our proposed methods are transfer learning since the additional features, which are extracted from the model pretrained on a larger dataset, provide direct or indirect knowledge transfer. Figure~\ref{fig: xumx} shows a high-level overview of the proposed training pipeline, indicating the use of VGGish features (purple) to modify the latent vector (red).  

\amy{Our approaches show some similarity to feature-based knowledge distillation approaches used in teacher-student learning \cite{gou2021knowledge}, where the pre-trained representations are used to regularize the embedding space during training.} VGGish features might contain information that is useful for separation but not adequately represented in the unregularized latent space. Projecting the latent vectors into VGGish feature space can help transfer the knowledge from VGGish feature space into latent vectors. Moreover, VGGish features have strong discriminative power. Forcing the latent vectors to be close to VGGish features can lead to more separable latent space representations, and prevent the model from confusing between distinct instruments.

\subsection{Source separation model}
We adopt the X-UMX model as the baseline model~\cite{sawata2020all} since this model achieves very good results on the MUSDB18 dataset and has open-sourced code \footnote{\href{https://github.com/sony/ai-research-code/tree/master/x-umx}{https://github.com/sony/ai-research-code/tree/master/x-umx}}. The model is based on the Open-Unmix (UMX)  architecture~\cite{stoter19}. Instead of training one separate model for each instrument, X-UMX uses a bridging networks architecture, connecting the paths to cross each source’s network by adding two average operators to the original UMX model. The result shows \unit[0.4]{dB} improvement for the average Source-to-Distortion Ratio (SDR) compared to baseline UMX architecture~\cite{stoter19}. As indicated in Figure~\ref{fig: xumx}, the model uses four separate encoders consisting of fully-connected layers and bi-directional recurrent layers to compute the latent vectors $l \in \mathbb{R}^{B\times T_l}$ where $B$ denotes the number of feature bins for the latent vector for each instrument \amy{while $T_l$ denotes the number of the latent vectors across time}. The decoders comprise of fully-connected layers to decode the target spectrogram masks from the latent vectors.

\subsection{Input representation and features}
Following \amy{the setup of the X-UMX model}, the input of the source separation model is the  magnitude spectrogram $x \in \mathbb{R}^{T_s\times F}$, where $T_s$ represents the duration of the spectrogram while $F$ represents number of frequency bins. The short time Fourier transform is computed with a hop length and block length of \unit[1024]  and \unit[4096], respectively. 

The additional VGGish features are extracted by the pre-trained VGGish model~\cite{45611} as \unit[128]-dimensional vector with \unit[0.96]{s} time resolution (no overlap). The features are PCA transformed (with whitening) and quantized to 8-bits. 

One obstacle encountered when incorporating the VGGish features is the difference in time resolution. The \amy{VGGish model is originally trained for clip-level tagging so the features have a lower resolution (approx.~\unit[0.96]{s}). Our source separation model needs frame-wise prediction so has a higher resolution (approx.~\unit[0.02]{s})}. We solve the problem of different time resolutions by simply repeating the features $n$ times with $n$ representing the number of time frames in \unit[0.96]{s}. This approach is based on the assumption that one frame of VGGish features contains information for the entire $n$ latent vector frames and that a slight mis-alignment in time \amy{will have a negligible effect on the results.} 

\subsection{Transfer Learning Approaches}
\begin{figure}
 \centerline{
 \includegraphics[width=0.8\columnwidth]{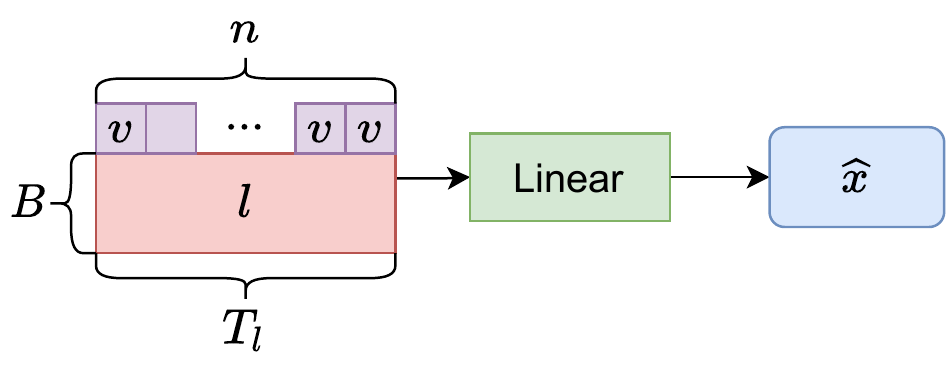}}
 \caption{The proposed \textit{Concat} method to concatenate VGGish features with latent vectors.}
 \label{fig: vgg-cat}
\end{figure}

\subsubsection{\amy{Method 1: Concatenation}}
The first proposed method aims to provide additional information encoded by VGGish for decoder to utilize during training. We simply concatenate the VGGish features (of the mixture audio) with the latent vector along feature dimension $B$, as shown in Figure~\ref{fig: vgg-cat}. It should be noted that this approach needs VGGish features during inference time. The original VGGish model has ~72M parameters. Although the VGGish features can be extracted prior to training so it would not influence training time too much, this approach has the potential to slow down the inference time.  Since in practice there is no access to the ground truth for the separated tracks during inference, the VGGish features we use in this method for both training and testing are extracted from the mixture audio. This method is referred as \textit{Concat} in the remainder of this paper.

\subsubsection{\amy{Method 2: Contrastive Regularization}}
The second method aims to regularize the latent space \amy{with the help of additional} VGGish features. To do so, we utilize the VGGish features extracted from the separate ground truth tracks (e.g., bass latent vectors from the model should be close to VGGish features extracted from the bass track) and add an extra loss term $L_\mathrm{con-reg}$ based on cosine similarity to force the latent vectors to be close to VGGish features: 
\begin{equation}\label{eq:vgg-close}
  L_\mathrm{con-reg} =
    \begin{cases}
      1-\cos(f(l), v) & \text{if $y=1$}\\
      \max(\cos(f(l), v)-\alpha,0) & \text{if $y=-1$} ,
    \end{cases}       
\end{equation}
with $y=1$ when the latent vector $l_i$ and the VGGish features $v_i$ both correspond to the same instrument $i$ (e.g., bass latent vector and bass VGGish features) and $y=-1$ in the case that $l_i$ and $v_j$ represent two different instruments $i \neq j$. The hyperparameter $\alpha$ is the margin of distance (set to $\alpha=0.2$  after hyperparameter search). A 1D CNN with \unit[1] kernel size ($f$) is applied to transform the latent vector dimensionality to $128$ to match the VGGish feature dimensionality, allowing us to compute the cosine distance between $l$ and $v$. This method is referred as \textit{Con-Reg} in the remainder of this paper. 

\begin{figure}
 \centerline{
 \includegraphics[width=0.8\columnwidth]{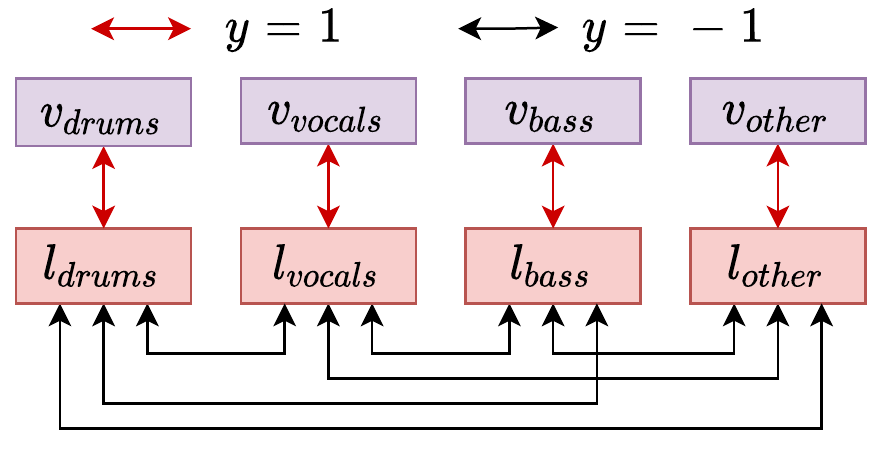}}
 \caption{The proposed \textit{Con-Reg} method to regularize latent vectors through VGGish features.}
 \label{fig: vgg-close}
\end{figure}

\subsubsection{\amy{Method 3: Distance-based Regularization}}
Similar to method \textit{Con-Reg}, the latent space is regularized with an additional loss. In this case, however, the additional loss term $L_\mathrm{dis-reg}$ aims at forcing the distances between pairs of latent vectors to be similar to the distances between two corresponding VGGish features:
\begin{equation}
  L_\mathrm{dis-reg} = \max\big(D_\mathrm{latent}(f(l_i), f(l_j)) - D_\mathrm{vgg}(v_i, v_j), 0\big) ,
\end{equation}
where $D_\mathrm{latent}$ represents the cosine distance between two latent vectors and $D_\mathrm{vgg}$ represents the distance of two VGGish feature vectors from two different instruments $i$ and $j$ where $i \neq j$. This method is similar to the scenario in \textit{Con-Reg} if $y=-1$ in Equation~(\ref{eq:vgg-close}). Instead of choosing a fixed value $\alpha$ as the margin, the distance of VGGish feature pairs serve as a `soft' margin which indicates the lower bound of the distance of two latent vectors. \amy{Since the cosine distance is evaluated on two distinct instruments, the latent vectors should be more separated from each other, indicating that $D_\mathrm{latent}$ should be smaller than $D_\mathrm{vgg}$} 

The loss design is based on the observation that the latent vectors from the same instrument category might not always be close to the same category. For example, `Electrical Bass' and `Double Bass' in `Bass' category might have slightly different features. The same as `Female Vocals' and `Male Vocals' in the `Vocals' categories. \amy{Since the VGGish model is pre-trained on the dataset that contains more detailed instrument labels\footnote{The ontology of the labels is shown in this website \href{https://research.google.com/audioset/ontology/index.html}{https://research.google.com/audioset/ontology/index.html} }, the distance of VGGish features might be able to capture the similarity of each instrument frame.} As a result, instead of using the category labels $y$ to determine the absolute distance, we use distance between the corresponding VGGish features to decide the distance of the latent vectors. Since optimizing the model to predict latent vectors matching the distance of VGGish features is not easy, we use the VGGish distance ($D_{\mathrm{vgg}}$) as a margin so that the latent vectors will have a distance ($D_{\mathrm{latent}}$) where $D_{\mathrm{latent}} \leq D_{\mathrm{vgg}}$. Distances smaller than $D_{\mathrm{vgg}}$ are also acceptable, since these encourages the latent vectors of two separate instruments to be more separable.  This method is referred as \textit{Dis-Reg} in the remainder of this paper. 

\subsection{Training Setup}
The Adam optimizer~\cite{kingma2014adam} is used with a 0.0001 learning rate and 0.00001 weight decay to optimize the model. Early stopping is applied if the validation loss does not decrease for 25 epochs and the learning rate decrease by a factor of 0.3 if the validation loss does not decrease for 10 epochs. 
The applied data augmentation (e.g., channel swapping and volume adjustment) is identical to the one described in~\cite{stoter19}. Each input sample has a length of \unit[6]{s} and is picked by randomly selecting a starting time in the audio. The standard Mean Square Error (MSE) loss and regularization loss (for \textit{Con-Reg} \& \textit{Con-Dis}) are combined with:
\begin{equation}
L_\mathrm{total} = L_\mathrm{mse} + \lambda L_\mathrm{reg},
\end{equation}
where $L_\mathrm{reg}$ is either $L_\mathrm{con-reg}$ or $L_\mathrm{dis-reg}$. The weight of the regularization loss $\lambda$ is set  after hyperparameter search to $\lambda_\mathrm{con-reg}=0.000001$ and $\lambda_\mathrm{dis-reg}=1$, respectively. Since $L_\mathrm{con-reg}$ has a `strong' margin between the latent vector and VGGish features, we found that setting the loss small to gradually influence the model during training can lead to better performance.  Note that the multi-domain loss in the X-UMX model is ignored in this study even if it led to an improvement of \unit[0.2]{dB}~\cite{sawata2020all} for the sake of computational speed.

\begin{figure*}
     \centering
     \begin{subfigure}[b]{.98\columnwidth}
         \centering
         \includegraphics[width=\textwidth]{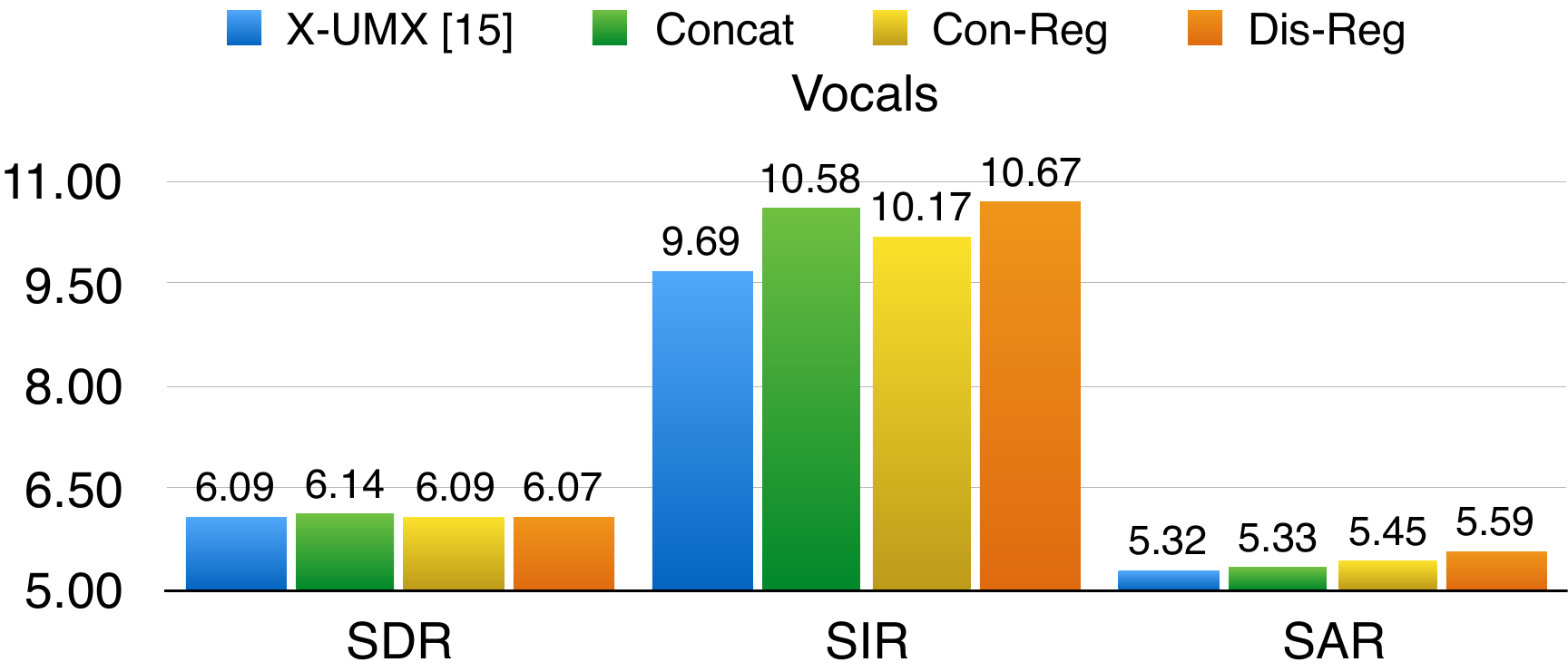}
     \end{subfigure}
     \hspace{3em}%
     \begin{subfigure}[b]{.98\columnwidth}
         \centering
         \includegraphics[width=\textwidth]{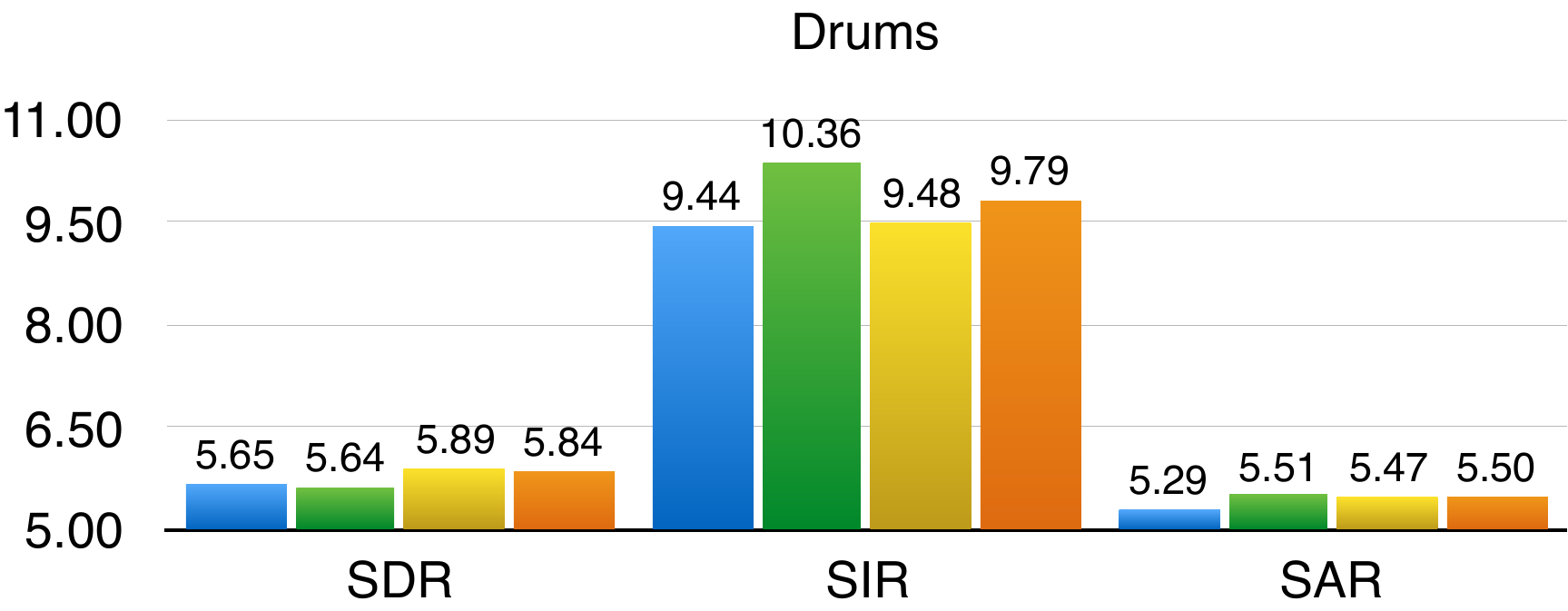}
     \end{subfigure}
     
     \vspace{2em}
     \begin{subfigure}[b]{.98\columnwidth}
         \centering
         \includegraphics[width=\textwidth]{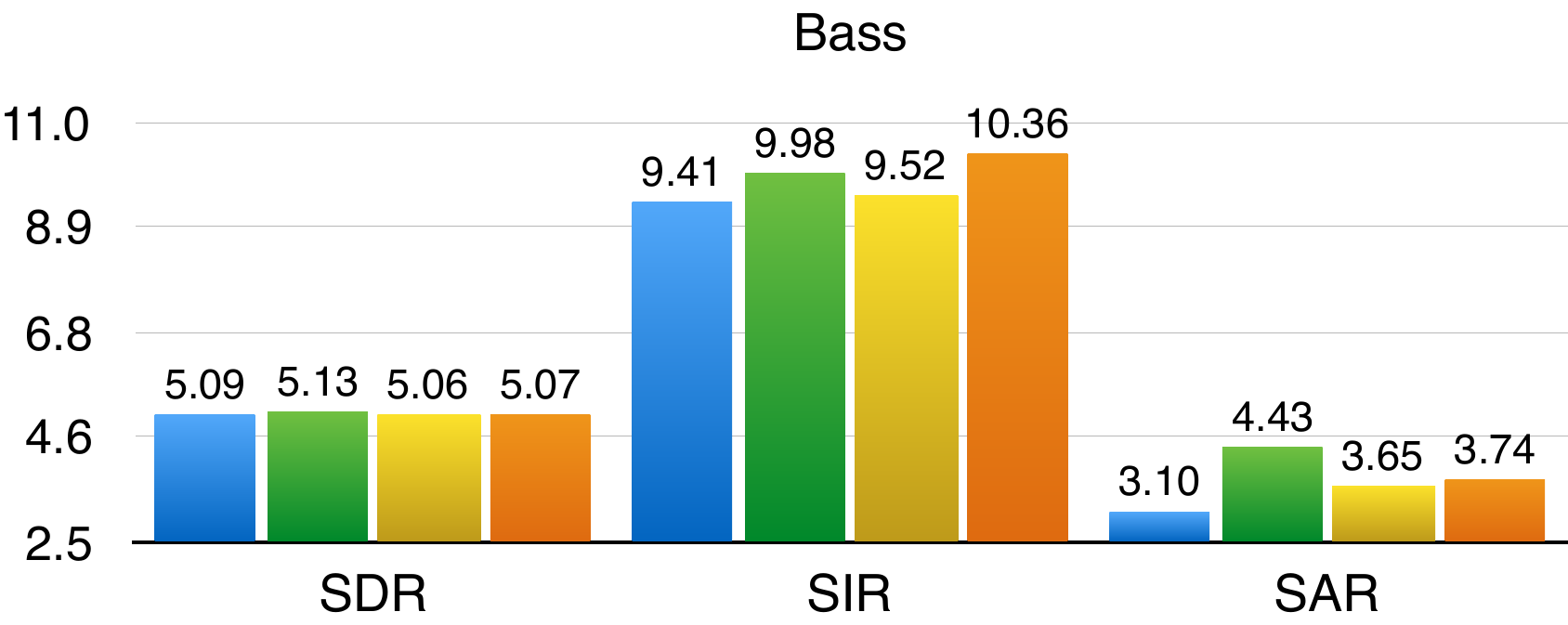}
     \end{subfigure}
     \hspace{3em}%
     \begin{subfigure}[b]{.98\columnwidth}
         \centering
         \includegraphics[width=\textwidth]{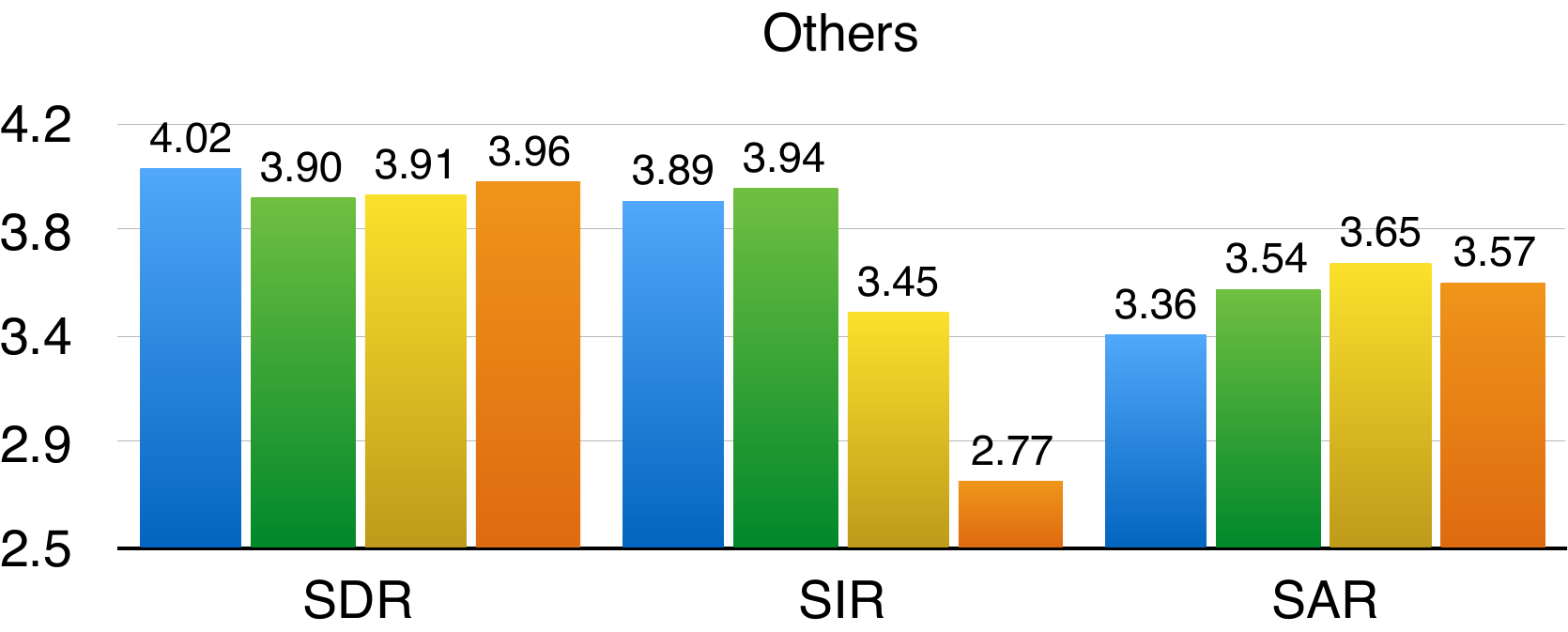}
     \end{subfigure}
        \caption{SDR, SIR and SAR scores for X-UMX and our proposed methods across four instruments.}
        \label{fig: result}
\end{figure*}

\begin{figure}[!b]
\centering
    \begin{subfigure}[b]{0.23\textwidth}
         \includegraphics[width=\textwidth]{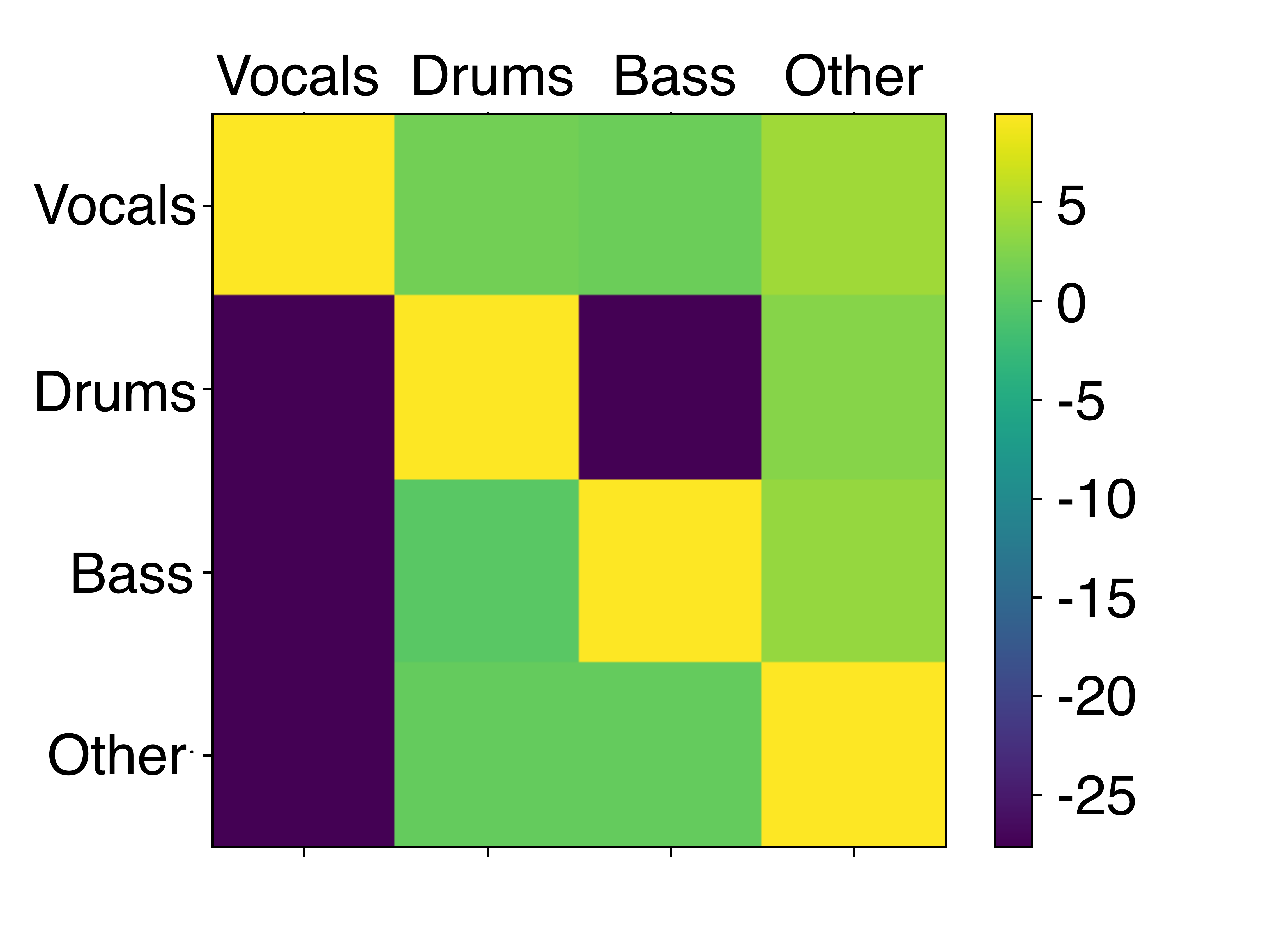}
         \caption{X-UMX}
    \end{subfigure}
    \hfill
    \begin{subfigure}[b]{0.23\textwidth}
         \includegraphics[width=\textwidth]{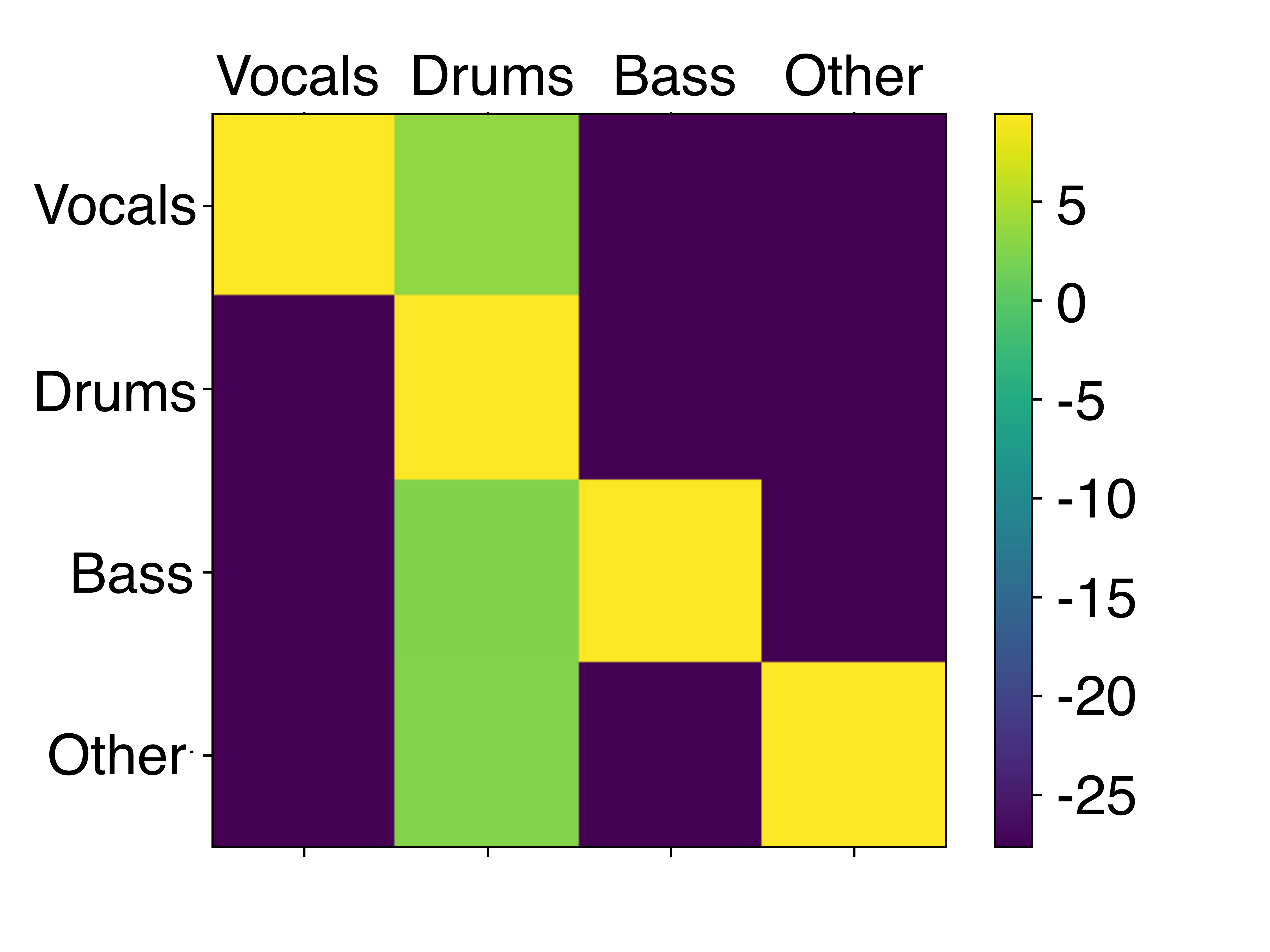}
         \caption{Concat}
     \end{subfigure}
     \hfill
     \begin{subfigure}[b]{0.23\textwidth}
         \includegraphics[width=\textwidth]{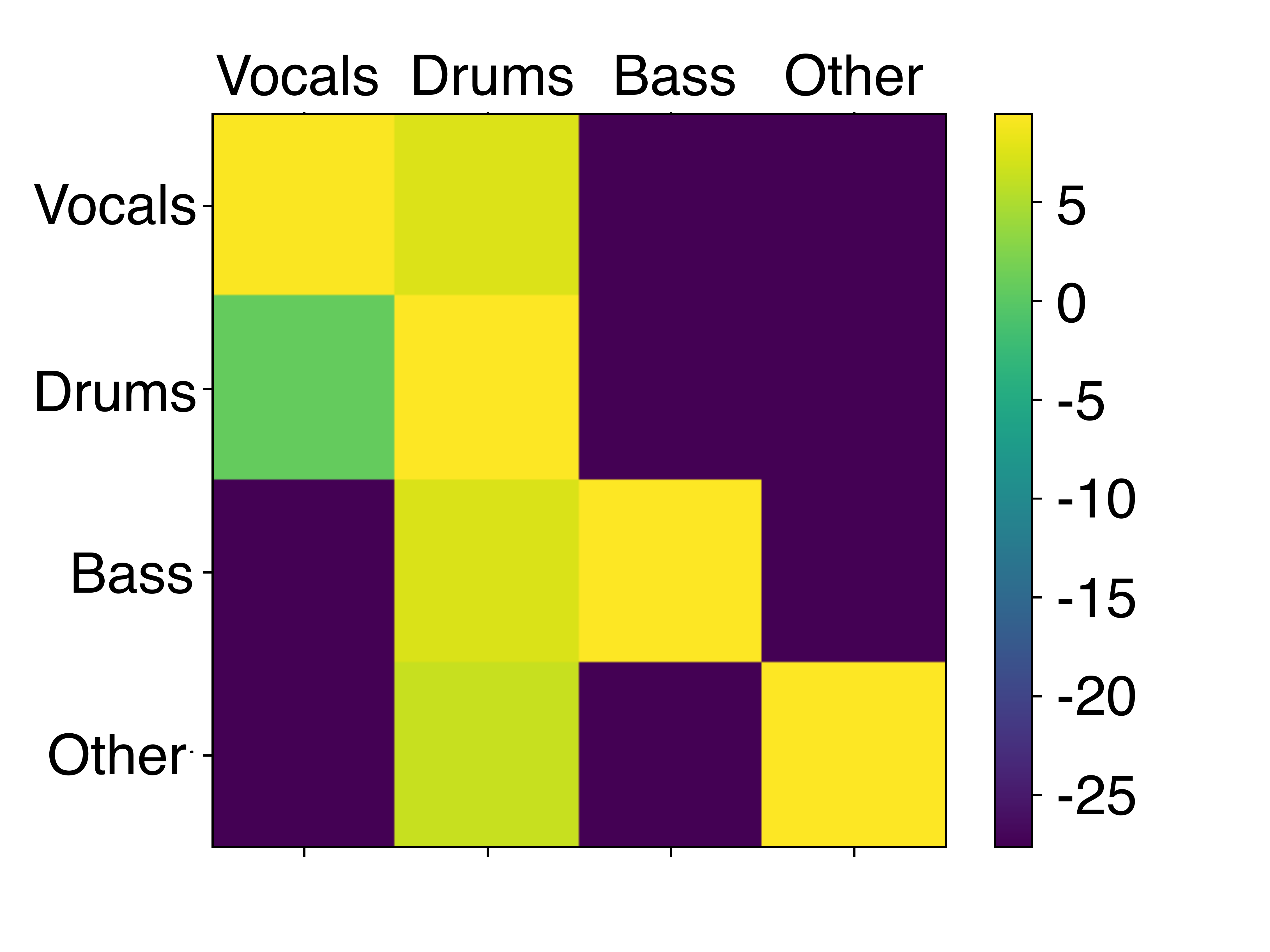}
         \caption{Con-Reg}
     \end{subfigure}
     \hfill
     \begin{subfigure}[b]{0.23\textwidth}
         \includegraphics[width=\textwidth]{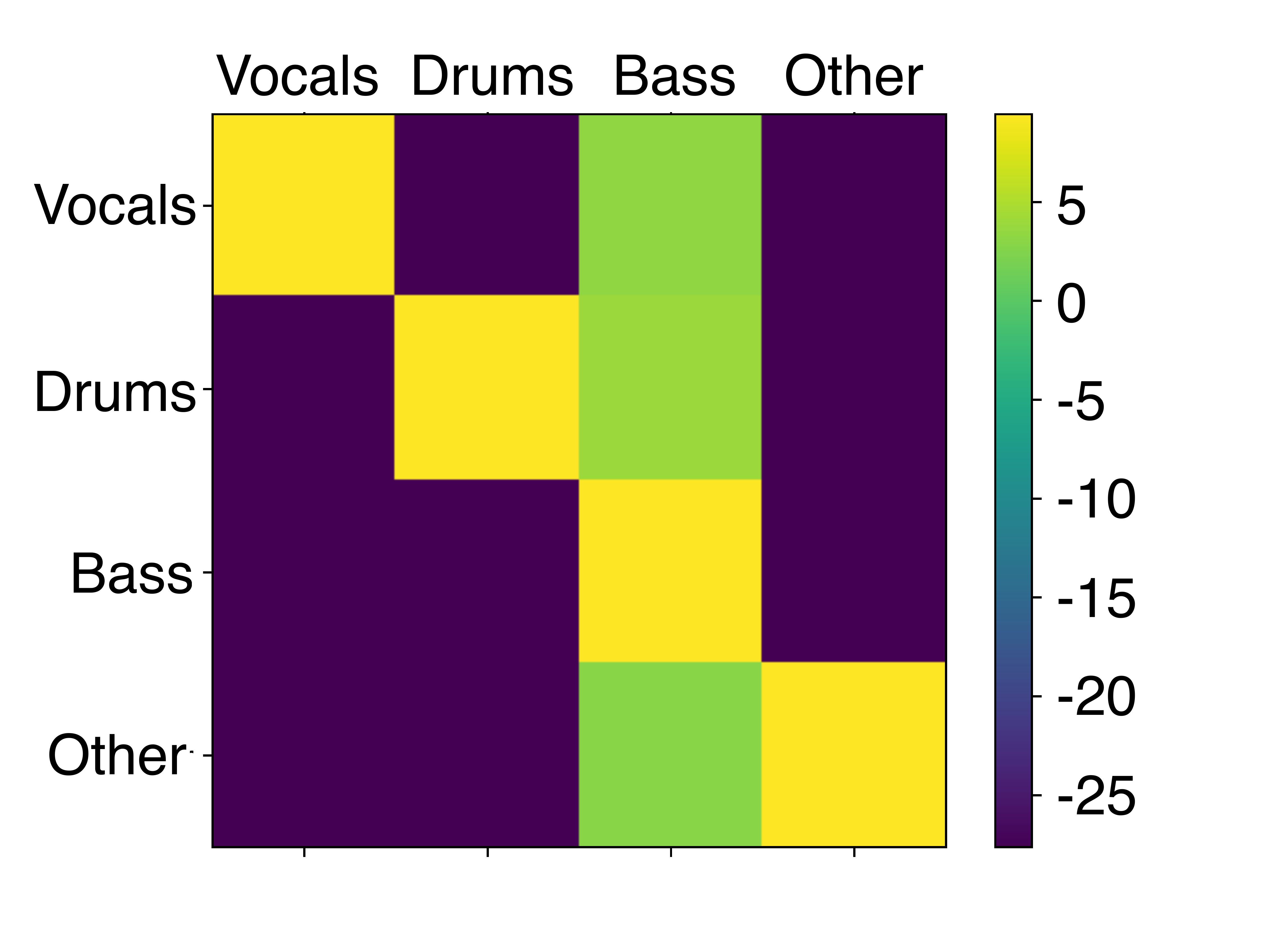}
         \caption{Dis-Reg}
     \end{subfigure}
 \caption{Confusion matrix of latent vector classification. }
 \label{fig: confusion}
\end{figure}

\section{Experiments} \label{sec: experiment}
We train and evaluate the proposed methods on the MUSDB18 dataset~\cite{musdb18} and use the `train,' `evaluation,' and `test' split defined in the original dataset. \amy{MUSDB18 has a total of 150 full-track songs of different styles. All signals are stereophonic and encoded at 44.1kHz. Each song in the dataset is comprised of four tracks, `Vocals,' `Bass,' `Drums,' and `Other.'} The X-UMX model without any regularization is our baseline for comparison. The same training strategy described above is also used to re-train the X-UMX. The \texttt{Museval} toolbox is used to calculate evaluation metrics: Signal-to-Distortion Ratio (SDR), Signal-to-Interference Ratio (SIR), and Signal-to-Artifact Ratio (SAR)~\cite{SiSEC18}. \amy{These three metrics are commonly used to evaluate the quality, the amount of other sources, and the amount of unwanted artifacts in an estimated source. Increasing values indicate better performance.}

\subsection{Music Source Separation}
The results of the source separation experiments per instrument are given in Figure~\ref{fig: result}. We can observe that compared to the baseline X-UMX model, \amy{our proposed methods can suppress unwanted artifacts and increase the SAR scores on all instruments. }The result is aligned with our assumption that VGGish features contain additional information which is helpful for separation. Utilizing VGGish features can help stabilize the model and producing fewer artifacts. 

\begin{figure*}
\centering
     \begin{subfigure}[b]{0.3\textwidth}
        \centering
         \includegraphics[width=\textwidth]{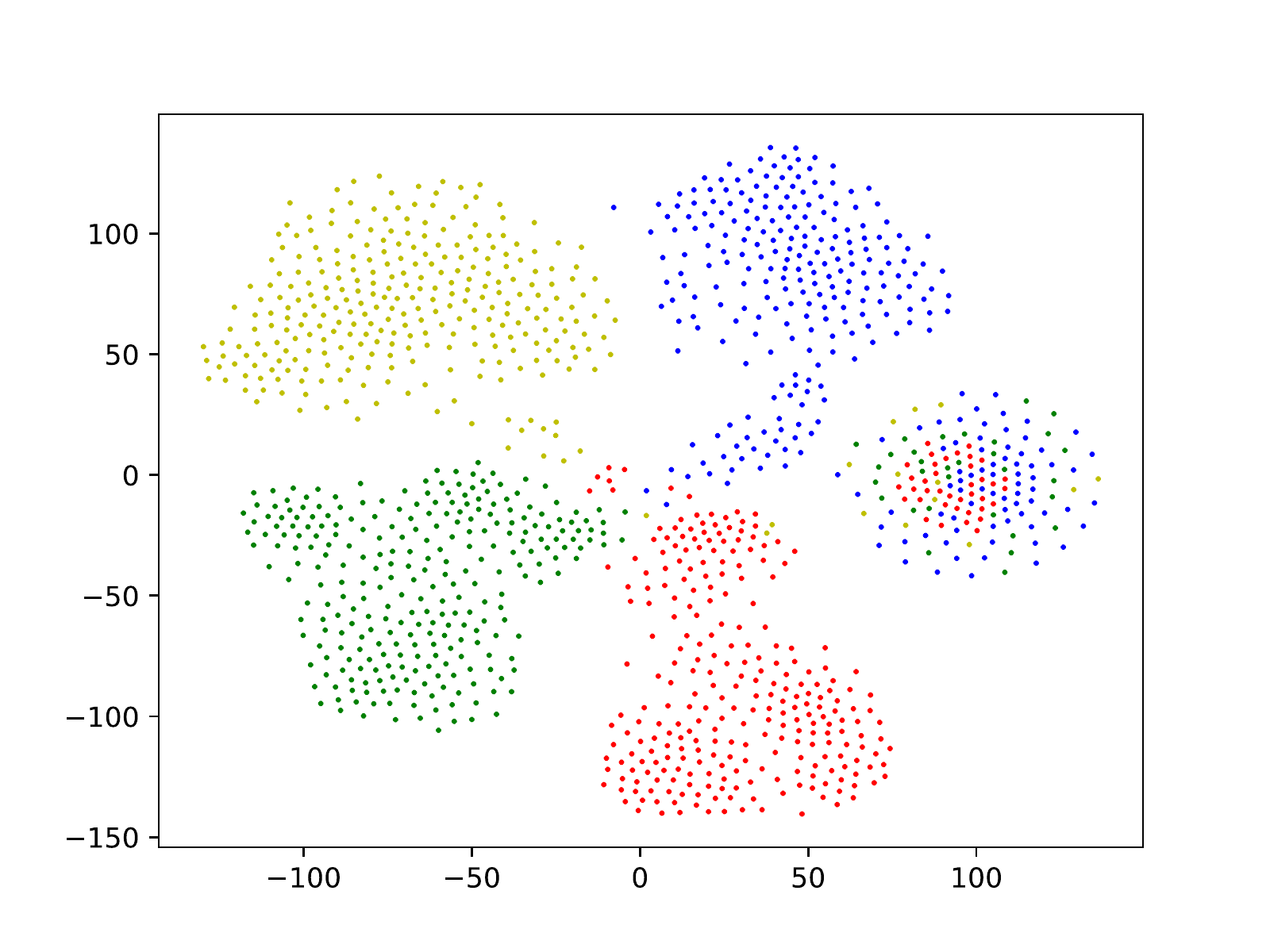}
         \caption{VGGish features}
         \label{fig: VGG-39}
     \end{subfigure}
     \begin{subfigure}[b]{0.3\textwidth}
        \centering
         \includegraphics[width=\textwidth]{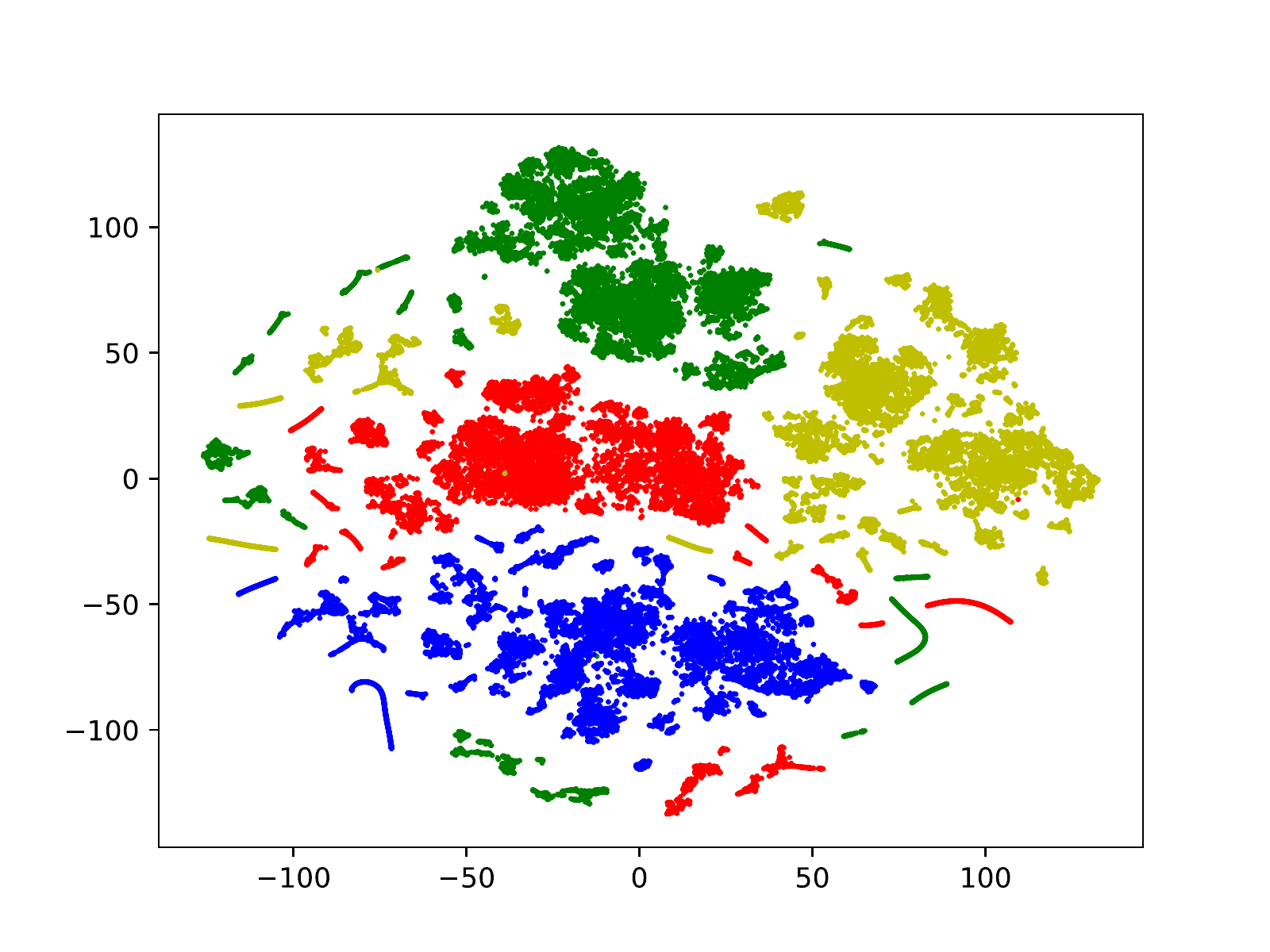}
         \caption{X-UMX}
         \label{fig: umx-39}
     \end{subfigure}
     \begin{subfigure}[b]{0.3\textwidth}
        \centering
         \includegraphics[width=\textwidth]{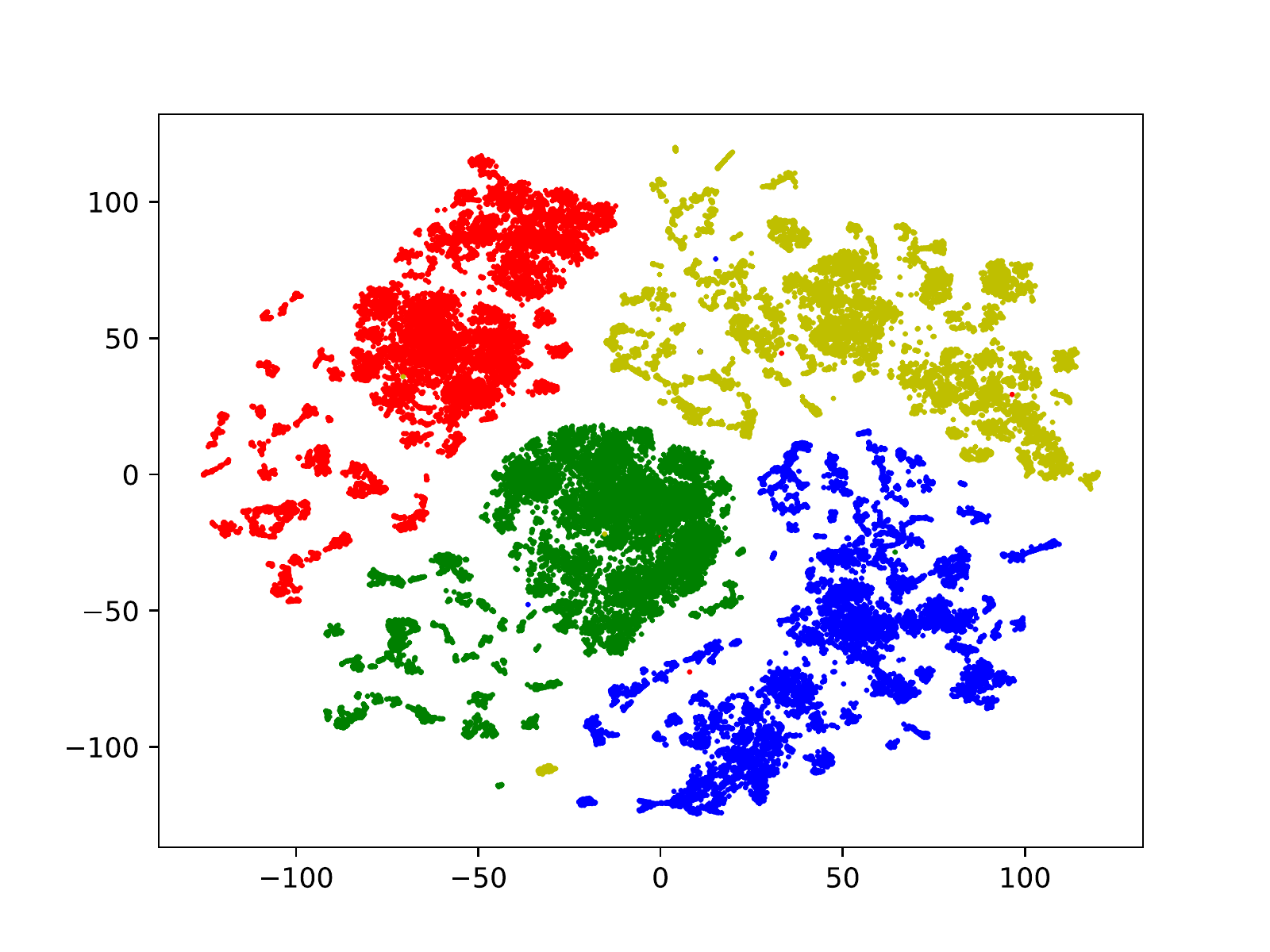}
         \caption{Concat}
         \label{fig: VGG-cat-39}
     \end{subfigure}
     \medskip
     \begin{subfigure}[b]{0.3\textwidth}
        \centering
         \includegraphics[width=\textwidth]{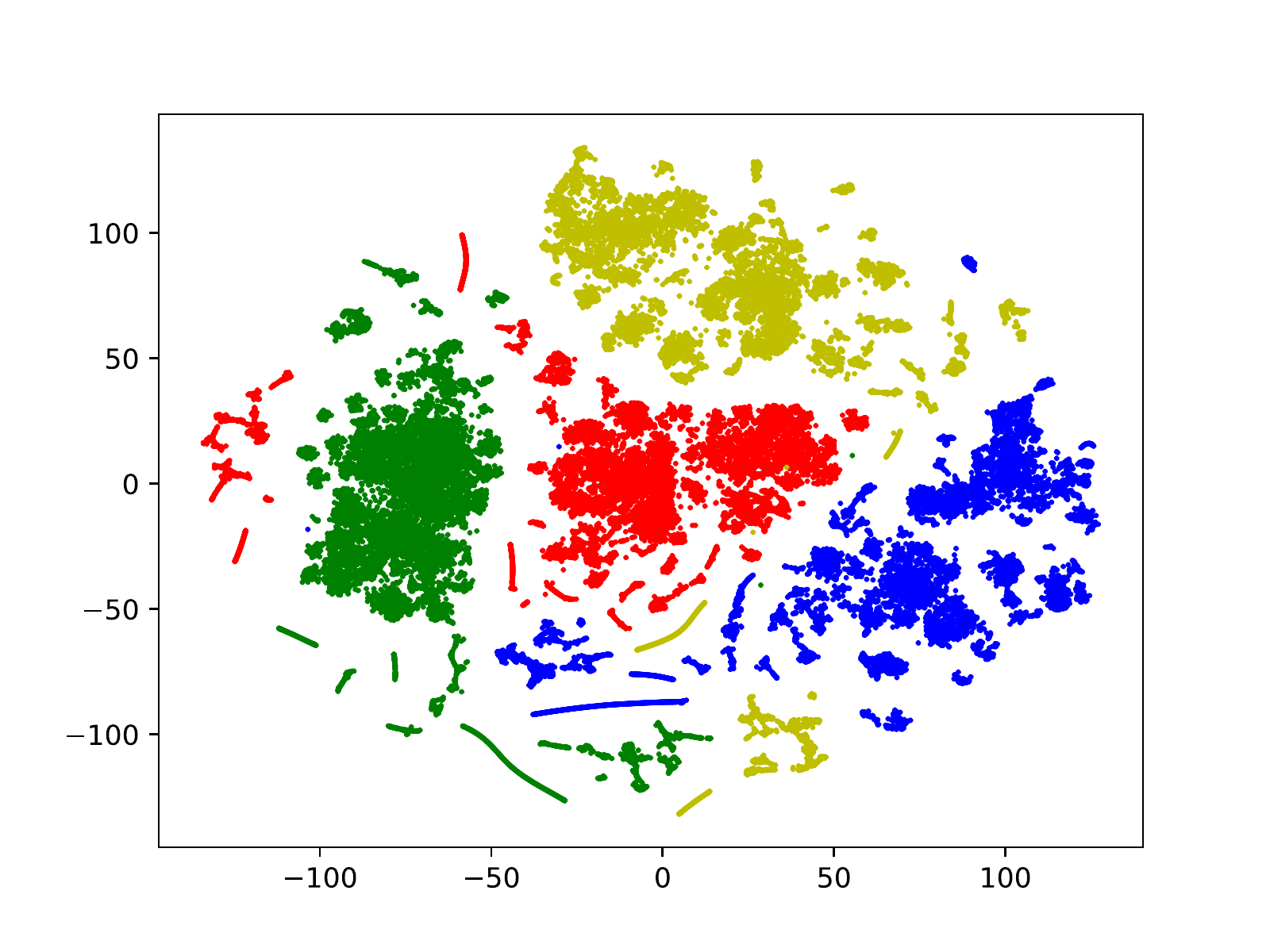}
         \caption{Con-Reg}
         \label{fig: VGG-close-39}
     \end{subfigure}
     \begin{subfigure}[b]{0.3\textwidth}
        \centering
         \includegraphics[width=\textwidth]{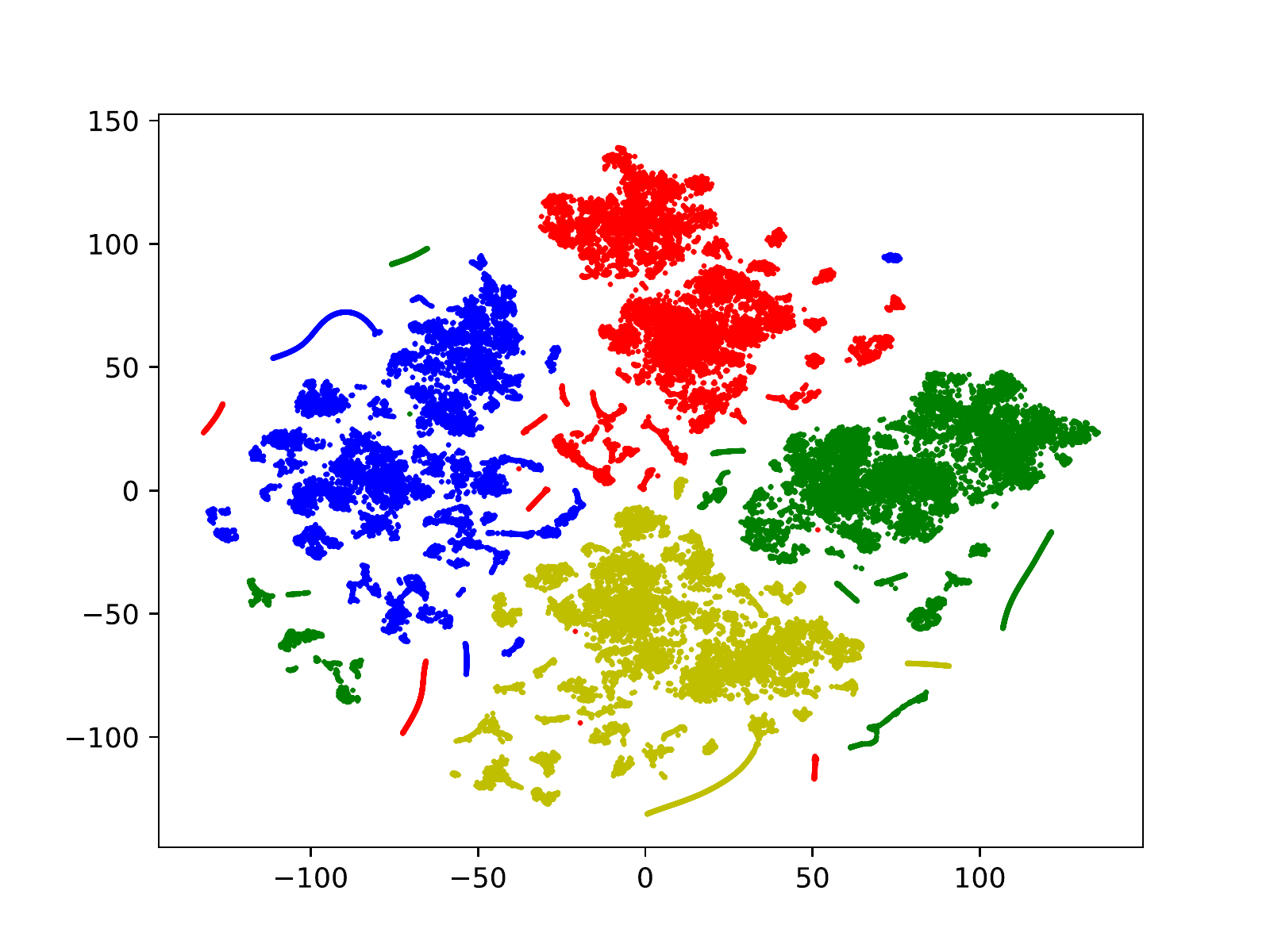}
         \caption{Dis-Reg}
         \label{fig: VGG-dis-39}
     \end{subfigure}
     
        \caption{The t-SNE visualization of the VGGish features as well as the latent vectors of the baseline X-UMX model and the proposed methods from an audio sample in test set. `Blue': vocals, `Red': drums, `Green': bass, `Yellow': other.}
        \label{fig:tsne}
\end{figure*}

By using the proposed methods, SIR scores also improve for `Vocals', `Drums' and `Bass'. \amy{This result suggests that the discriminative power of the VGGish features can benefit separation and decrease the interference of non-targeted sources.} As our proposed methods aim to make the latent space more discriminative by either adding additional VGGish features or forcing different instruments' latent vector to be apart from each other, the interference from other instruments decrease. The soft margin employed in \textit{Dis-Reg} generally leads to a higher SIR score than the hard margin in \textit{Con-Reg}. The SIR score of `Other,' however, decreases after regularization. Since `Other' includes a variety of instruments, the distance of each latent vector might vary substantially. For example, if `Other' contains low pitched instruments such as Tuba, then the distance might be closer to `Bass' instead of `Violin', which is also from `Other'. As a result, the distance regularization might actually impede the training and lead to a poor SIR score. The result also suggests the possibility of improving source separation system further if the `Other' category is replaced by additional ``clean'' instruments.

Comparing \textit{con-Reg} and \textit{Dis-Reg}, we can observe that generally \textit{Dis-Reg} outperforms \textit{Con-Reg}. The distance-based regularization helps the model maintain a structure targeted at source separation while adding relevant information. The SDR score stays mostly constant for most of the instruments except for `Drums,' where we can observe improvement when using regularization methods. Since VGGish features are a compact representation of the audio signal, they lack the detail needed to improve the quality of the separated sound.  

\amy{To summarize, while the \textit{Concat} method performs better than \textit{Dis-Reg} and \textit{Con-Reg} on most of the metrics, it needs to incorporate the VGGish model during inference, which increases computation time and required memory. In contrast, \textit{Dis-Reg} and \textit{Con-Reg} although do not achieve similar improvement as \textit{Concat} on some of the metrics, they still lead to improvement compared to X-UMX and do not need additional features during inference.}


\subsection{Latent Space Classification}
To measure the discriminative power of the latent vectors after proposed latent space modification, we compute the latent vectors from the test set for our four models: X-UMX, \textit{Concat}, \textit{Con-Reg}, and \textit{Dis-Reg}. Each audio input has a $T_l \times C$ latent features for $C$ represents the dimension of latent vectors after modification. Since all the tracks have same length, the total number of frames across classes is equally distributed. For \textit{Con-Reg} and \textit{Dis-Reg}, $C$ is equal to \unit[128]. $C$ is equal to ($B$+128) for \textit{Concat} and $C$ is equal to $B$ for X-UMX. We perform K-means algorithm on all the latent vectors across all audio inputs for clustering. The latent vectors from the same instruments should be in the same cluster. 


The confusion matrix of the proposed models are shown in Figure~\ref{fig: confusion}. The matrix is processed through log function for better visualization. `Vocals' has the best performance for X-UMX latent vectors, while `Other' tends to be confused with other instruments. The confusion is eased by using latent vectors from proposed methods. Surprisingly, when using \textit{Concat} and \textit{Con-Reg}, `Drums' tends to get confused with other instruments and `Bass' tends to to get confused when using \textit{Dis-Reg}. The consistent confusion of a specific instrument might be caused by the silence or low volume frames. K-means algorithm directly assigns the closest instrument to those frames.

\subsection{Latent Space Visualization}

We randomly choose a sample from the test set and visualize a t-SNE projection~\cite{van2008visualizing} of the extracted latent vectors from the VGGish features, the baseline X-UMX model, \textit{Concat}, \textit{Dis-Reg}, and \textit{Con-Reg}. The results are shown in Figure~\ref{fig:tsne}. We can observe from Figure~\ref{fig: umx-39} that~---although the latent vectors are clearly clustered---~several small clusters can be found to be far from their corresponding main cluster (top-left yellow and bottom-right red). \amy{This implies that model might easily confuse one instrument with another.} When using regularization in Figure~\ref{fig: VGG-close-39} and Figure~\ref{fig: VGG-dis-39}, the clusters tend to be more tightly packed and the spurious small clusters decrease in both number and size. Directly concatenating the VGGish features with latent vectors yields the best separable latent space, as shown in Figure~\ref{fig: VGG-cat-39}. As occurs with classification, we observe that some of the clusters separated from the main clusters are from low volume audio frames or silence. Since they contain minimum instrument information, latent vectors are difficult to cluster into one of the four groups. Moreover, we can observe that \textit{Concat} and \textit{Dis-Reg} achieve better clustering results than X-UMX and \textit{Con-Reg}. The result is aligned with our separation result in Figure~\ref{fig: result}, where \textit{Dis-Reg} and \textit{Concat} achieve the highest improvement.

\section{Conclusion}
In this work, we propose three methods to incorporate VGGish features into SOTA music source separation system. \amy{The first method simply concatenates features with the latent vectors, while the other two novel methods regularize the latent space}. Compared to the baseline X-UMX model, our proposed methods can reduce the artifacts and the interference created by the model and improve the SAR and SIR scores on `Vocals', `Drums' and `Bass'. Latent space visualization shows that the latent space has better discriminative power after regularization. 
The proposed methods can be easily incorporated into other model architectures and adopted to other features. Therefore, we plan to include other features into our proposed training strategies, such as the L3-Net embedding features~\cite{cramer2019look}. Separation of more instruments will also be studied in the future to fully leverage the discriminative power of VGGish features. Finally, applying our presented regularization to other audio-related tasks will be explored in the future.

\nocite{*}
\bibliographystyle{IEEEbib}
\bibliography{DAFx22_tmpl} 
\end{document}